\documentclass{article}
\usepackage{spconf,amsmath,graphicx}
\usepackage{graphics} % for pdf, bitmapped graphics files
\usepackage{epsfig} % for postscript graphics files
\usepackage{mathptmx} % assumes new font selection scheme installed
\usepackage{times} % assumes new font selection scheme installed
\usepackage{amssymb}  % assumes amsmath package installed
\usepackage{multirow}
\usepackage{subfigure}
\usepackage{siunitx}
\usepackage[table,xcdraw]{xcolor}
\usepackage{url}
\usepackage{adjustbox}
\usepackage[table,xcdraw]{xcolor}
\usepackage{tabularx}
\usepackage{makecell}
\usepackage{cellspace}
\usepackage{pgfplots}
\pgfplotsset{width=5.5cm,compat=1.5}
\title{ORTHONORMAL EMBEDDING-BASED DEEP CLUSTERING \\FOR SINGLE-CHANNEL SPEECH SEPARATION}

\name{Soyeon Choe$^{1}$, Soo-Whan Chung$^{1,2}$,
Youna Ji$^{2}$, and Hong-Goo Kang$^{1}$}
\address{$^{1}$Department of Electrical and Electronic Engineering,
Yonsei University, Seoul, South Korea\\
$^{2}$Naver Corporation, Seongnam-si, Gyeonggi-do, South Korea}
\begin{document}
%\ninept
%
\maketitle
%

%%%%%%%%%%%%%%%%%%%%%%%%%%%%%%%%%%%%%%%%%%%%%%%%%%%%%%%%%%%%%%%%%%%%%%%%%%%%%%%%
\begin{abstract}
 Deep clustering is a deep neural network-based speech separation algorithm that first trains the mixed component of signals with high-dimensional embeddings, and then uses a clustering algorithm to separate each mixture of sources. In this paper, we extend the baseline criterion of deep clustering with an additional regularization term to further improve the overall performance.  
 This term plays a role in assigning a condition to the embeddings such that it gives less correlation to each embedding dimension, leading to better decomposition of the spectral bins. 
 The regularization term helps to mitigate the unavoidable permutation problem in the conventional deep clustering method, which enables to bring better clustering through the formation of optimal embeddings. We evaluate the results by varying embedding dimension, signal-to-interference ratio (SIR), and gender dependency. The performance comparison with the source separation measurement metric, \textit{i.e.} signal-to-distortion ratio (SDR), confirms that the proposed method outperforms the conventional deep clustering method.

\end{abstract}

% \vspace{-12pt}

 %%%%%%%%%%%%%%%%%%%%%%%%%%%%%%%%%%%%%%%%%%%%%%%%%%%%%%%%%%%%%%%%%%%%%%%%%%%%%%%%
% \vspace{-10pt}
\section{introduction}
Speech signals in real environment often involve distortions such as noise, interfering speakers, and reverberation~\cite{gannot}. It is still a challenging task to enhance speech in the interfering speaker conditions although the conventional  speech enhancement algorithms obtain reasonable performance when the background noise is stationary~\cite{visser2003icassp}. Therefore, the need for separating mixed signals into number of sources through speech separation arises~\cite{ikram}. Speech separation technique can be used to solve the cocktail party problem~\cite{qian2018,cherry1953} in automatic speech recognition (ASR) systems by enhancing the speech of the specific speaker among multiple speakers along with various background noises.

One of the speech separation techniques is computational auditory scene analysis (CASA)~\cite{cooke1993,hu2}, which studies how humans separate speech and learn from them. However, this technique needs to manually design ad-hoc rules due to limited number of observations \cite{hu2004}. Another technique is non-negative matrix factorization (NMF)~\cite{smaragdis2007aslp, simsekli2014icassp}, which uses hand-designed rules from human observations with the assumption that audio spectrogram has a low rank structure that can be represented with a small number of bases~\cite{lee2000}. Even though this technique seems successful, the speaker dependency issue of the basis led to ad-hoc constraints, and there are also limitations in real-time applications due to the complexity of the decomposition process caused by large number of basis~\cite{lee2000}. 

With the increasing popularity of deep learning applications in various fields, how to make good use of deep learning technique in speech separation has become an essential factor nowadays. Hershey et al.~\cite{hershey2016icassp} proposed a deep clustering-based speech separation framework with embedding matrix used as a key factor to cluster the mixed signals, and the neural network is trained to obtain embedding vectors corresponding to each element of high-dimensional signal. 
Since embeddings are represented in time-frequency (T-F) domain and each input signal can be reconstructed by clustering the embedding vectors, it is possible to separate all the sources independently. 
Even though deep clustering is somehow helpful to overcome the permutation problem, each separated output still contains undesired segments if there is an error during the clustering process. Therefore, further deep learning-based speech separation frameworks, \textit{i.e.} deep attractor network (DANet)~\cite{chen2017icassp} and permutation invariant training (PIT)~\cite{yu2017icassp, kolbaek2017taslp}, were released to solve this issue. However, the permutation problem is still inevitable if the performance of constructed embedding or clustering is not high enough.

Our proposed work aims to develop an effective deep clustering framework by designing the training strategy to assign an additional regularization condition. 
By training the network to find the minimum loss between the embeddings and the identity matrix, the network automatically assigns an orthonormality constraint to the embeddings. Since each embedding can be considered as a basis of each input speaker, the proposed training method leads to have more embedding independency so that embeddings are useful for better decomposition. This effective formation of embeddings helps to improve the remaining permutation problem of the conventional method as the decomposed spectral bins help increase the performance of clustering. We demonstrate improvements of the proposed algorithm over the conventional deep clustering algorithm, and also analyze the results in various experimental setups.

% \vspace{-10pt}
\section{Deep clustering}
 
 Deep clustering is a deep network that uses learned feature transformations known as embeddings, to separate speech on open set of speakers~\cite{hershey2016icassp}.  In audio signals, raw input signal $x$ can be defined as a feature vector $X_{t,f}=g_{t,f}(x)$, where $t$ and $f$ represent frame and frequency index of the signal. Deep clustering starts with the assumption that a reasonable partition of elements exists for each region. Its objective is to find that partition using embedding matrix and K-means clustering in order to estimate masks to be applied to each input mixture of $X$. With audio signals, these regions are sets of time-frequency bins where each source dominates the other source. 
 Deep clustering seeks K-dimensional embeddings of the mixed signal to apply simple clustering in embedding space. 
 
 In training stage, deep neural network is used to transform input $x$ to K-dimensional embeddings $V\in \mathbb{R}^{\left \{ N \times K \right \}}$, with the embedding considered as unit-norm, $\left | v_{i} \right |^{2} = 1$, where $v_{i}$ is the embedding for element $i$. Embeddings $V$ actually represent \textit{N} by \textit{N} estimated affinity matrix, which analyzes the content of each mixed signal in terms of a set of channels. The affinity matrix $VV^{T}$ is learned to match $YY^{T}$ by minimizing the cost using the squared Frobenius norm as follows:
 \begin{equation}
    C_{Y}(V) = \left \| VV^{T} - YY^{T} \right \|_{F}^{2}.
 \end{equation}
  $YY^{T}$ is the target affinity matrix which is generated with the label indicator, $Y$, \textit{i.e.} the concept of comparing each T-F bin and assigning value '1' to the dominant bin, formed as an ideal binary mask (IBM). Note that low-rank formulation leads to efficient implementation of the cost function as shown in the equation below:
 \begin{equation}\label{eq:dc}
     C_{Y}(V)=\left \| V^{T}V \right \|_{F}^{2} - 2\left \| V^{T}Y \right \|_{F}^{2}+\left \| Y^{T}Y \right \|_{F}^{2}.
 \end{equation}
 
In inference stage, K-means clustering is used to cluster the K-dimensional embeddings into the number of sources. By assigning the value '1' to the spectral bin of each cluster and '0' to the other, ideal binary masks are formed for each source, and applying these masks to the mixed signal separates the signal into individual sources. Therefore, training to obtain appropriate embedding matrix is essential to cluster the embeddings sufficiently. 

    \begin{figure}[!t]
      \centering
      \includegraphics[width=83mm, clip=true]{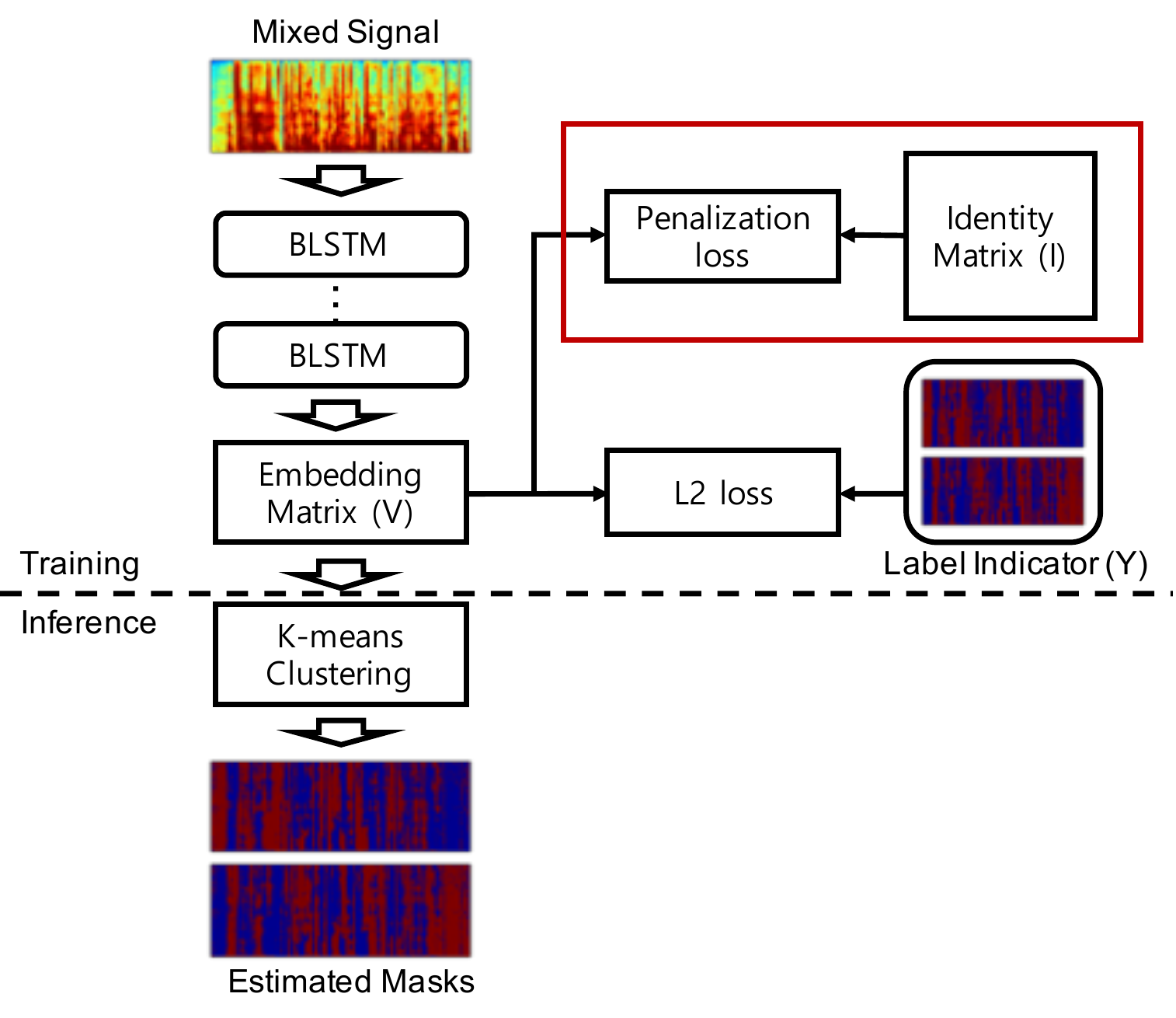}
%  \vspace{-8pt}     
      \caption{Proposed system architecture}
      \label{overall}
%  \vspace{-8pt}
  \end{figure} 

This proposed work has focused on enabling better clustering through the formation of optimal embeddings. The work of Hershey et al.~\cite{hershey2016icassp} considers learning the embedding to follow the label indicator, however, it contains lack of regularity. Therefore, the present study trains with more regularization to generate orthonormal embedding.

% \vspace{-10pt}
\section{Orthonormality of embedding matrix}
Even though the deep clustering method obtains high performance in single-channel speech separation tasks, there are still more improvements to be made. Knowing that the formation of optimal embeddings leads to better decomposition of spectral bins, focusing on regularizing the embeddings would be effective to the performance.

\vspace{-6pt}
\subsection{Penalization term}
\vspace{-2pt}
Penalization term is one of the regularization terms to encourage the diversity in annotation vectors as indicated below:
\begin{equation}\label{eq:pen}
P = \left \| V^{T}V - I \right \|_{F}^{2},
\end{equation}
where \textit{V, I}, and $\left \| \cdot \right \|_{F}$ are embeddings, identity matrix, and Frobenius norm, respectively. Computing $V^{T}V$, sets the output dimension to become embedding dimension by embedding dimension ($\mathbb{R}^{K \times K}$), so that it is easier to compute the loss between the identity matrix, $\mathbb{R}^{K \times K}$,  and $V^{T}V$. This term forces the diagonal elements of the embedding matrix to 1 and the off-diagonal to 0, making the embedding matrix to be orthonormal. Applying this term on deep clustering embeddings will lead to assign more independence to each other, resulting in efficient clustering. Like the efficient implementation of the cost function of deep clustering in Equation \eqref{eq:dc}, Equation \eqref{eq:pen} can be shown as the equation below:
 \begin{equation}
     P=\left \| VV^{T} \right \|_{F}^{2} - 2\left \| V \right \|_{F}^{2}+\left \| I \right \|_{F}.
 \end{equation}
 Therefore, the overall training cost function of the proposed system is as follows:

  \begin{equation}\label{eq:overall}
  \vspace{-3pt}
  \begin{split}
         C_{Y}(V)=(\left \| V^{T}V \right \|_{F}^{2} - 2\left \| V^{T}Y \right \|_{F}^{2}+\left \| Y^{T}Y \right \|_{F}^{2})\\
         + (\left \| VV^{T} \right \|_{F}^{2} - 2\left \| V \right \|_{F}^{2}+\left \| I \right \|_{F}),
  \end{split}
 \end{equation}
 which helps learn the embeddings to match the target affinity matrix and be orthonormal, and the whole system architecture is shown in Fig.~\ref{overall}. A similar idea of the penalization term has been introduced in a self-attentive word embedding task~\cite{lin2017iclr} without being used for any other types of signals.
 
 \vspace{-6pt}
  \subsection{Training procedure}
  \vspace{-2pt}
In order to train the proposed algorithm with a deep learning network, the ideal binary mask (IBM) was used as the target, \textit{i.e.} power dominant bin was set to '1', and '0' for others in each time-frequency bin. We used two bi-directional long short-term memory (BLSTM) layers followed by one fully connected layer. Each BLSTM layer consisted of 512 hidden cells with the size of the embedding dimension $K$. Adam optimizer with the learning rate of $10^{-4}$ was used for training, and the hyperbolic tangent function was used for both activation and output function of the network. The training criteria were L2 loss between affinity matrices and the additional penalization term, and the dropout rate was set to 0.5 for regularization. Also, we investigated several network models by varying embedding dimensions to 20, 30, and 40. 

 \vspace{-6pt}
 \subsection{Visualization of embedding covariance matrix}
 \vspace{-2pt}

\begin{figure}[!t]
      \centering
      \includegraphics[width=90mm, clip=true]{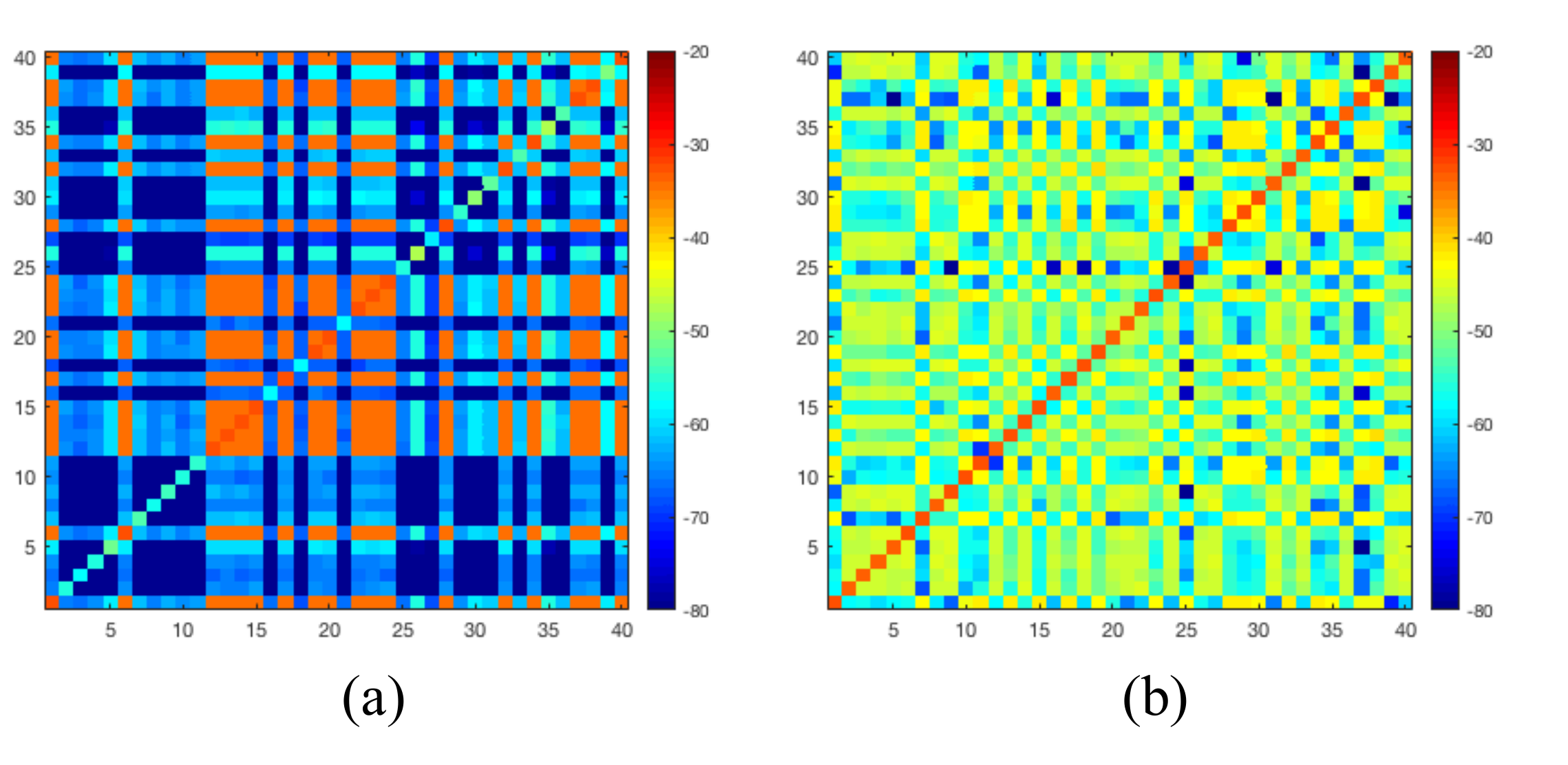}
%  \vspace{-8pt}     
 \caption{(a) Covariance matrix for conventional deep clustering method. (b) Covariance matrix for proposed method. The embedding dimension is set to 40.}
      \label{cov}
%  \vspace{-8pt}
  \end{figure}

Existing speech separation techniques, \textit{i.e.} deep clustering~\cite{hershey2016icassp}, DAN~\cite{chen2017icassp}, PIT~\cite{yu2017icassp, kolbaek2017taslp}, simply decompose the label indicator through the projection without any established standards. Therefore, it is difficult to figure out the degree of decomposition. The penalization term in our proposed method is a criterion that enables to sparsely project the label indicator to the embedding matrix, and it can be verified with Fig. \ref{cov}. This figure indicates the covariance matrix among embedding dimensions of 40, and it compares that of both conventional deep clustering and our proposed method. Fig. \ref{cov}(a) demonstrates that the conventional deep clustering method has some focused data on specific embedding dimensions; whereas, our proposed method equally spreads the data throughout all dimensions as shown in Fig. \ref{cov}(b).

% \vspace{-2pt}
% \vspace{-8pt}
\section{Speech Separation Experiments}
\vspace{-6pt}
\subsection{Experimental setup}
\vspace{-2pt}
We used Wall Street Journal (WSJ) corpus, both WSJ0 and WSJ1~\cite{paul1992} to make speech mixtures. We randomly select 10000 utterances for training and 5000 utterances for both validation and evaluation sets, where the same proportion of females and males is chosen. Especially for the evaluation set, a set of 1000 mixed utterances of 2 gender cases, \textit{i.e.} same gender and different gender mixtures, is used with interfering ratio of 3, 6, 9, 12, and 15 dB. Since SIR is evenly distributed in the evaluation set, it becomes easier to analyze the mixtures among gender and SIR. Sampling frequency of all data is set to 16 kHz, and the input feature $X$ is log spectral magnitude using STFT with 512 samples and 256 samples window shift of hanning window. Different from the conventional deep clustering network, our network uses whole utterance for the input instead of 100 frames, and the deep neural network is implemented with the Tensorflow libraries~\cite{tensorflow}. 

\vspace{-6pt}
\subsection{Speech separation procedure}
\vspace{-2pt}
At inference stage, speech separation is performed by K-means clustering. The output mask is generated by clustering the embeddings $V$, \textit{i.e.} the output from the proposed model for each utterance. The number of clusters is set to 2 since we assume to have 2 speaker mixtures in this experiment\footnote{It is possible to extend the system to 3 or more speaker mixtures.}. Among various types of masks, we generated binary masks with the clustered outputs to apply them to the mixture and obtain separated sources. We evaluated two types of deep clustering models, \textit{i.e.} the conventional deep clustering model and our proposed model, and compared the results. For choosing the target speech from the clustered output, we chose the speech with dominant power level as the target speech and the other as the interfering speech. For all experiments, averaged signal-to-distortion ratio (SDR) was used as the evaluation metric with \textit{mir\_eval} library \cite{eval2014ismir}.

\vspace{-3pt}
\section{Results and Discussion}
We tried to analyze the proposed results in three cases: SDR vs. dimension, SDR vs. SIR, and SDR vs. gender. With the first case, we wanted to see the effect of the penalization term among different embedding dimensions. As shown in Table~\ref{sir}, adding penalization criterion output better result with 20 embedding dimension than the baseline with 40 embedding dimension. Also, as the embedding dimension increased, the improvement rate between proposed method and the baseline increased, meaning that the signals decompose better as embedding dimension increases. From this experiment, we confirmed that adding penalization term to the deep clustering method assigns less correlation to the embeddings, resulting in improvements in performance. That is because less correlated embeddings result in better decomposition of spectral bins and lead to preferable clustering.

With the second case, we wanted to see the effect of penalization term among different SIRs. Table~\ref{sir} also shows the absolute SDR in embedding dimension 20 and 40 for various SIRs. As SIR increased, the effect of penalization term increased; however, there were some points that we needed to focus on. In embedding dimension of 20, the penalization term gave negative effect with SIR of 3 dB, meaning that the embedding dimension below 20 may be not enough to handle low SIR mixtures. Forcing to give orthonormality to embeddings can break the correlation when it is actually needed; therefore, it could give more distortion to the signal instead. 

\begin{table}[!t]\setlength\tabcolsep{5pt}
% \vspace{-5pt}
\caption{SDR vs. SIR}
\vspace{5pt}
\label{sir}
\begin{tabular}{cc|ccccc|c}
\hline
                          &         & \multicolumn{5}{c}{\textbf{SIR (dB)}}                      \\ \hline
\multicolumn{1}{c|}{\textbf{Dim.}} & \textbf{Meth.}  & \textbf{3}      & \textbf{6}     & \textbf{9}     & \textbf{12}     & \textbf{15}     & \textbf{Avg.} \\ \hline
\multicolumn{1}{c|}{}     & DC      & 3.57  & 6.30 & 9.05 & 12.01 & 14.91 & 9.17   \\ \cline{2-8} 
\multicolumn{1}{c|}{40}   & Prop.   & 3.60  & 6.66 & 9.18 & 12.82 & 15.35 & 9.64   \\ \cline{2-8} 
\multicolumn{1}{c|}{}     & Imprv. & 0.03  & 0.36 & 0.13 & 0.81  & 0.44  & \textbf{0.47}   \\ \hline
\multicolumn{1}{c|}{}     & DC      & 3.51  & 6.35 & 9.18 & 11.85 & 14.53 & 9.10   \\ \cline{2-8} 
\multicolumn{1}{c|}{20}   & Prop.   & 3.50  & 6.56 & 9.56 & 12.36 & 15.08 & 9.49   \\ \cline{2-8} 
\multicolumn{1}{c|}{}     & Imprv. & -0.01 & 0.21 & 0.38 & 0.51  & 0.55  & \textbf{0.39}   \\ \hline
\end{tabular}
\end{table}
% \vspace{5pt}

\begin{table}[!t]\setlength\tabcolsep{5pt}
% \vspace{-5pt}
\caption{SDR vs. Genders}
\vspace{5pt}
\label{gender}
\centering
\begin{tabular}{ccccccc}
\hline
                               & \multicolumn{3}{c}{\textbf{Same Gender}}              & \multicolumn{3}{c}{\textbf{Mixed Gender}} \\ \hline
\multicolumn{1}{c|}{\textbf{Dim.}} & \textbf{DC}    & \textbf{Prop.} & \multicolumn{1}{c|}{\textbf{Imprv.}} & \textbf{DC}       & \textbf{Prop.}     & \textbf{Imprv.}   \\ \hline
\multicolumn{1}{c|}{40}        & 9.07 & 9.20 & \multicolumn{1}{c|}{0.13}   & 9.27    & 10.09    & \textbf{0.82}     \\ \hline
\multicolumn{1}{c|}{30}        & 9.12 & 9.09 & \multicolumn{1}{c|}{-0.03}  & 9.05    & 9.79     & \textbf{0.74}     \\ \hline
\multicolumn{1}{c|}{20}        & 9.05 & 8.96 & \multicolumn{1}{c|}{-0.09}  & 9.16    & 9.86     & \textbf{0.70}     \\ \hline
\end{tabular}
\end{table}

For the last case, we wanted to analyze the result between genders, and we found that the penalization term works better on mixtures with different genders than same genders as shown in Table~\ref{gender}. With embedding dimension of 20, the same gender mixtures showed degradation in performance with penalization term. This confirmed that the embedding dimension below 20 is not enough to increase the effect of penalization term. Also, we observed that the penalization term works better with embeddings that already have some independence, such as the mixture signals with mixed gender. 
% Fig. \ref{tsne} in section 3 confirmed that the mixed gender mixtures has distinct characteristics within each gender, leading to have less correlation in the embeddings.

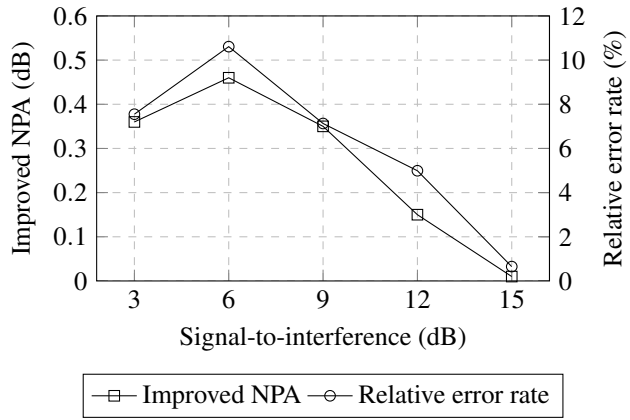
\begin{figure}[t]
    \centering
    \begin{tikzpicture}
    \pgfplotsset{set layers}
    \begin{axis}[
        scale only axis,
        width=0.7*\linewidth,height=\linewidth*0.41,
        xlabel={Signal-to-interference (dB)},
        ylabel={Improved NPA (dB)},
        ymin=0, ymax=0.6,
        xtick={3,6,9,12,15},
        xticklabels={3,6,9,12,15},
        ytick={0,0.1,0.2,0.3,0.4,0.5,0.6},
        y tick label style={
             /pgf/number format/.cd,
            fixed,
            % fixed zerofill,
            precision=2,
            /tikz/.cd
        },
        ymajorgrids=true,
        xmajorgrids=true, 
        grid style=dashed
    ]
    \addplot+[
        color=black,
        mark=square,
        ]
        coordinates {
        (3, 0.36)(6,0.46)(9,0.35)(12,0.15)(15,0.01)
        };\label{plot_one}
        % \addlegendentry{Improved NPA (dB)}
    \end{axis}
    \begin{axis}[
        scale only axis,
        width=0.7*\linewidth,height=\linewidth*0.41,
        ylabel={Relative error rate (\%)},
        axis y line*=right,
        axis x line=none,
        ymin=0, ymax=12,
        xtick={3,6,9,12,15},
        xticklabels={3,6,9,12,15},
        ytick={0,2,4,6,8,10,12},
        legend columns=2,
        legend style={at={(0.5,-0.35)},anchor=north},
        grid style=dashed   
    ]
    \addlegendimage{/pgfplots/refstyle=plot_one}\addlegendentry{Improved NPA}
    \addplot+[
        color=black,
        mark=o,
        ]
        coordinates {
        (3,7.55)(6,10.61)(9,7.13)(12,4.99)(15,0.65)
        };\label{plot_two}
    \addlegendentry{Relative error rate}
    \end{axis}
    \end{tikzpicture}
    \caption{Quality measurements of estimated IBMs}
 \label{ibm}
  \vspace{-2pt}
\end{figure}

Also, with the analysis of improved normalized projection alignment (NPA) metric~\cite{schmid} and relative error rate, we have verified that the permutation problem of the conventional deep clustering method has been improved with our proposed system, as shown in Fig. \ref{ibm}. Improved NPA shows the improvement in mask formation between deep clustering and our proposed method, and the relative error rate compares the improved bin prediction error between the target and estimated masks. As the undesired segments in the separated outputs indicate the permutation problem, the quality of the estimated masks is associated with the effective separation of mixed speech. With high SIR mixtures, conventional deep clustering method already obtains high performance, meaning it has less permutation problems. On the other hand, low SIR mixtures has difficulty in forming an optimal mask in speech separation and need more improvements. Therefore, Fig.~\ref{ibm} shows that both improved NPA and relative error rate generally increase as the SIR decreases.

\vspace{-5pt}
\section{Conclusion}
In this paper, we have observed the efficacy of our proposed method through the comparison with the conventional deep clustering method. Analyzing the covariance matrix of the embedding output of both conventional and prosed methods, we have confirmed that penalization term sparsely projects the label indicator to all the dimensions in the embedding matrix. Through diverse experiments, we have verified that the performance of our proposed method is maximized with high embedding dimension, high SIR, and different gender, which already contains some independence assumption between speakers in the mixture. Also we have improved the permutation problem in the conventional deep clustering method, and it is verified with the evaluation metric for the estimated masks. 
Through the verification of the efficacy of penalization term on embeddings, we are to further extend our studies by implementing this term on other embedding-based speech separation methods, \textit{e.g.} deep attractor network \cite{chen2017icassp}. Also, we will analyze the potentials and limitations of our method and stabilize the penalization term to optimize its effect on the embeddings. 

\noindent\textbf{Acknowledgements.} 
This work was supported and funded by Clova AI, Naver corporation.

% \bibliographystyle{unsrt}
%\bibliography{ieeeproc.bib}
\vfill
\newpage
\label{sec:refs}
% \begin{thebibliography}{}
%   \input{ieeeprocshort.bib}
% \end{thebibliography}
\bibliographystyle{IEEEbib}
\bibliography{ieeeprocshort}
% \vspace{-5pt}

\end{document}